# Distributed order Hausdorff derivative diffusion model to characterize non-Fickian diffusion in porous media

Yingjie Liang[*], Wen Chen, Wei Xu, HongGuang Sun[*]

Institute of Soft Matter Mechanics, College of Mechanics and Materials, Hohai University, Nanjing, Jiangsu 211100, China

**Corresponding authors**: Yingjie Liang liangyj@hhu.edu.cn

HongGuang Sun shg@hhu.edu.cn

*Abstract*: Many theoretical and experimental results show that solute transport in heterogeneous porous media exhibits multi-scaling behaviors. To describe such non-Fickian diffusions, this work provides a distributed order Hausdorff diffusion model to describe the tracer transport in porous media. This model is proved to be equivalent with the diffusion equation model with a nonlinear time dependent diffusion coefficient. In conjunction with the structural derivative, its mean squared displacement (MSD) of the tracer particles is explicitly derived as a dilogarithm function when the weight function of the order distribution is a linear function of the time derivative order $p(\alpha)=2c\alpha$. This model can capture both accelerating and decelerating anomalous and ultraslow diffusions by varying the weight parameter *c*. In this study, the tracer transport in water-filled pore spaces of two-dimensional Euclidean is demonstrated as a decelerating sub-diffusion, and can well be described by the distributed order Hausdorff diffusion model with $c = 1.73$. While the Hausdorff diffusion model with $\alpha = 0.97$ can accurately fit the sub-diffusion experimental data of the tracer transport in the pore-solid prefractal porous media.

# 1. Introduction

Contaminant transport in porous media is one of the most important topics in environmental fluid mechanics [1], such as radionuclide transport through fractured rock [2], seawater invasion in karstic [3], coastal aquifers [4] and contaminant migrates at waste landfill site on fractured porous media [5]. Understanding the solute transport can help to control the contaminant diffusion, remediate polluted water and provide ways to sustainably use the natural resources [6]. It is well known that diffusion of solute particles in free space is Fickian diffusion and satisfies the popular Einstein relationship [7], i.e., $<x^2(t)> \sim t$, where $<x^2(t)>$ is the mean squared displacement (MSD). The MSD provides a common denominator to classify the diffusion processes in heterogeneous porous media [8]. However, many theoretical, experimental and field results show that the Einstein relationship is not valid for particles diffusion in heterogeneous porous media. The MSD becomes a nonlinear function, such as power law form used to describe the anomalous diffusion [9], the logarithmic form [10] and the inverse Mittag-Leffler form used to characterize the ultraslow diffusion [11], i.e., superslow diffusion [12] and more complicated cases with more than one exponent to describe multi-scaling behaviors [13-14].

To describe such non-Fickian diffusions in complex media, several mathematical and physical models were successively proposed from the perspective of non-integral derivative order, such as the fractional and Hausdorff derivative diffusion models [15-16], the variable order fractional and Hausdorff derivative diffusion models [14, 17], the distributed fractional derivative diffusion model [18-19] and the structural derivative diffusion model [11, 20]. The fractional and the Hausdorff derivative diffusion models describe the anomalous diffusion with mono-scaling behavior by connecting the derivative orders with structure changes in the media. The variable order fractional and Hausdorff derivative diffusion models, respectively generalized the fractional and the Hausdorff derivative diffusion models, can well capture the Non-Fickian dynamics with multi-scaling behavior. In both of the models, the derivative orders are time and space dependent ones, which reflect the isotropic feature of the heterogeneous media. The distributed order fractional derivative diffusion model is also a kind of popular models to depict multi-scale non-Fickian transport dynamics, and its order is a time or space derivative order independent function [19]. The structural derivative diffusion model is proposed to explore the physical mechanism of the ultraslow diffusion, which diffuses even more slowly than the sub-diffusion [11]. The structural function used in the definition of the structural derivative diffusion equation determines the ultraslow dynamics evolution. Importantly, the variable order and distributed order models mentioned above can describe the multi-scaling behaviors. However, the distributed order Hausdorff derivative diffusion model has not been reported, and its physical mechanism and ability to capture the multi-scaling behaviors needs to be verified.

The Hausdorff derivative, i.e., fractal derivative, does not contain convolution and is a local one compared with the fractional derivative [15, 21]. The Hausdorff derivative is defined based on the Hausdorff fractal metrics in the Hausdorff fractal space [22], which has clear geometrical feature. The time derivative order in the Hausdorff derivative diffusion model quantifies the Hausdorff fractal dimension of the diffusion trajectory [23]. The fundamental solution of the Hausdorff derivative diffusion model is a stretched Gaussian distribution [22]. It should be pointed out that the Hausdorff derivative is a special case of the local structural derivative, when the structural function is a power law function [11]. In this study, the distributed order Hausdorff derivative model is proposed to describe the tracer transport in porous media. For a specific function of the order distribution, i.e., weight function, the MSD of the solute particles controlled by the distributed order of Hausdorff diffusion model is derived in conjunction with the local structural derivative. The corresponding structural function has specific relationship with the weight function. The proposed model is tested to investigate the solute diffusion in water-filled pore spaces of two-dimensional Euclidean and pore-solid prefractal porous media [24].

This paper is organized as follows. In Section 2, the methodologies used to derive the distributed order time Hausdorff diffusion equation are introduced. In Section 3, evolution patterns in the distributed order Hausdorff diffusion equation are detected and the distributed-order Hausdorff diffusion model is tested by two simple cases. The results are discussed in Section 4. Finally, in Section 5 we draw some conclusions.

## 2. Methodologies

### 2.1 Distributed order time fractional diffusion equation

The time fractional diffusion equation for the propagator $u(x,t)$ [9], i.e., the probability of finding the solute particle at position $x$ at time $t$,

$$\frac{\partial^\beta u(x,t)}{\partial t^\beta} = D\frac{\partial^2 u(x,t)}{\partial x^2}, \quad u(x,0) = \delta(x), \tag{1}$$

where $D$ is the diffusion coefficient, a positive constant, $\beta \in (0,1]$ the time fractional derivative order. In Eq. (1), the definition of the fractional derivative in the Caputo sense is used [25].

$$\frac{\partial^\beta u(x,t)}{\partial t^\beta} = \frac{1}{\Gamma(1-\beta)}\int_0^t d\tau (t-\tau)^{-\beta}\frac{\partial u(x,t)}{\partial t}. \tag{2}$$

In this scenario, the MSD is a power law function of time $<x^2(t)> = 2Dt^\beta$. This model describes the time-dependent diffusion process.

The distributed order time fractional diffusion equation is constructed based on Eqs. (1) and (2) [26].

$$\int_0^1 d\beta\, p(\beta)\frac{\partial^\beta u(x,t)}{\partial t^\beta} = D\frac{\partial^2 u(x,t)}{\partial x^2}, \quad u(x,0) = \delta(x), \tag{3}$$

where the weight function $p(\beta)$ is non-negative function of the time derivative order, and satisfies the following relationship [26]

$$\int_0^1 p(\beta)d\beta = c > 0. \tag{4}$$

$p(\beta)$ is considered the probability density of the derivative order when the positive constant $c$ is 1. The form of the weight function in Eq. (3) differentiates different types of solute transport.

Here we just show two cases of weight function [27], $p(\beta) = \delta(\beta-\beta_0)$, $0 < \beta_0 \le 1$, a Dirac delta function, and $p(\beta) = 1$, i.e., uniformly distributed order. When $p(\beta) = \delta(\beta-\beta_0)$, $0 < \beta_0 \le 1$, the MSD is derived as [27]

$$<x^2(t)> = \frac{2}{\Gamma(1+\beta_0)} D t^{\beta_0}, \tag{5}$$

which can depict the Brownian motion when $\beta_0 = 1$, and when $\beta_0 < 1$, it is a sub-diffusion process.

When $p(\beta) = 1$, the explicit form of the MSD is [27]

$$<x^2(t)> = 2D(\ln t + \gamma + e^t E_1(t)), \tag{6}$$

where the Euler constant $\gamma = 0.5772$, and the exponential integral $E_1(t)$ is defined as

$$E_1(z) = \int_z^\infty \frac{e^{-t}}{t} dt. \tag{7}$$

The asymptotic expressions at small and large times can be easily obtained based on the properties of the above exponential integral in Eq. (7),

$$<x^2(t)> \sim \begin{cases} 2Dt \ln(1/t), & t \to 0 \\ 2D \ln(t), & t \to \infty \end{cases}. \tag{8}$$

It is a slightly super-diffusion at small times, but at lager times it is an ultraslow diffusion. Thus, the distributed order time fractional diffusion equation is capable to model multi-scale diffusion process by using specific weight function. More details can be found in Ref. [27].

**2.2 Hausdorff derivative and structural derivative diffusion equations**

The basic strategy of the time Hausdorff derivative is not complicated [23]. When considered a particle movement under fractal time, its movement distance is calculated by

$$s(t) = v(t - t_0)^\alpha, \tag{9}$$

where $s$ denotes the distance, $t$ the current time instance, $v$ the uniform velocity, $t_0$ the initial instance, $\alpha$ the fractal dimensionality in time. The Hausdorff integral distance can be derived when the velocity depends on time.

$$s(t)=\int_{t_0}^{t} v(\tau)d(\tau-t_0)^\alpha. \tag{10}$$

Based on Eq. (10), the time Hausdorff derivative is defined as,

$$\frac{ds}{dt^\alpha} = \lim_{t_1 \to t} \frac{s(t_1)-s(t)}{(t_1-t_0)^\alpha -(t-t_0)^\alpha} = \frac{1}{\alpha(t-t_0)^{\alpha-1}} \frac{ds}{dt}, \tag{11}$$

and when the initial instance $t_0 = 0$, Eq. (11) is reduced to

$$\frac{ds}{dt^\alpha} = \lim_{t_1 \to t} \frac{s(t_1)-s(t)}{t_1^\alpha - t^\alpha} = \frac{1}{\alpha t^{\alpha-1}} \frac{ds}{dt}. \tag{12}$$

By using the time Hausdorff derivative in Eq. (12), the corresponding diffusion equation is constructed, which is an alternative method to describe sub-diffusion

$$\frac{\partial u(x,t)}{\partial t^\alpha} = D\frac{\partial u(x,t)}{\partial x^2}, \quad u(x,0) = \delta(x) \tag{13}$$

where the time Hausdorff derivative order $\alpha \in (0,1]$. The Hausdorff derivative is local in time. Eq. (13) can be restated as a traditional diffusion equation by using the fractal metric in time $\hat{t} = t^\alpha$. The solution of Eq. (13) yields a stretched Gaussian distribution [15].

$$u(x,t) = \frac{1}{\sqrt{4\pi Dt^\alpha}} \exp\left(-\frac{x^2}{4Dt^\alpha}\right). \tag{14}$$

Its MSD is a power law function of time.

$$<x^2(t)> = 2Dt^\alpha. \tag{15}$$

In more complicated media, the fractal structure cannot well be described by the power law fractal metrics. The structural metrics is proposed to generalize the fractal metric [28]. In the context of structural metrics, Eqs. (9-10) are respectively extended to following forms,

$$s(t)=vk(t-t_0), \tag{16}$$

and

$$s(t)=\int_{t_0}^{t} v(\tau)dk(\tau-t_0),\tag{17}$$

where $k$ is the structural function. The local structural derivative is easily obtained,

$$\frac{ds}{d\,k(t)}=\lim_{t_1\to t}\frac{s(t_1)-s(t)}{k(t_1-t_0)-k(t-t_0)}.\tag{18}$$

The local structural derivative diffusion equation generalizes the Hausdorff derivative diffusion equation, which is flexible to classify diffusion processes in conjunction with the structural function

$$\frac{d\,u(x,t)}{d_k t}=D\frac{\partial^2 u(x,t)}{\partial x^2},\tag{19}$$

where $d_k$ represents the structural derivative. When $k(t)=t^\alpha$, it is the Hausdorff derivative with the initial instance $t_0 = 0$. The solution of Eq. (19) is a rescaled Gaussian distribution using the metric transform $\hat{t}=k(t)$.

$$u(x,t)=\frac{1}{\sqrt{4\pi Dk(t)}}\exp\left(-\frac{x^2}{4Dk(t)}\right).\tag{20}$$

When the structural function is an inverse Mittag-Leffler function $k(t)=E_\alpha^{-1}(t)$, $0 < \alpha \leq 1$, Eq. (20) describes the general ultraslow diffusion. The mean squared displacement is derived as

$$<x^2(t)>=2DE_\alpha^{-1}(t).\tag{21}$$

It should be pointed out that when $\alpha = 1$, $E_\alpha^{-1}(t)$ degenerates into the logarithmic function $\ln(t)$, which is correlated with the classical Sinai diffusion [29].

**2.3 Distributed order time Hausdorff diffusion equation**

Based on the Hausdorff derivative and the distributed order time fractional diffusion equation, the distributed order time Hausdorff diffusion equation is constructed

$$\int_0^1 d\alpha\, p(\alpha) \frac{\partial u(x,t)}{\partial t^\alpha} = D \frac{\partial^2 u(x,t)}{\partial x^2}, \quad u(x,0) = \delta(x), \tag{22}$$

where $\alpha \in (0,1]$ is the time Hausdorff derivative order, the weight function $p(\alpha)$ is non-negative function of the order,

$$\int_0^1 p(\alpha) d\alpha = c > 0. \tag{23}$$

The Hausdorff derivative is equivalent with the following form [30].

$$\frac{\partial u(x,t)}{\partial t^\alpha} = \frac{t^{1-\alpha}}{\alpha} \frac{\partial u(x,t)}{\partial t}. \tag{24}$$

Thus, Eq. (22) can be transformed into

$$\int_0^1 d\alpha\, p(\alpha) \frac{t^{1-\alpha}}{\alpha} \frac{\partial u(x,t)}{\partial t} = D \frac{\partial^2 u(x,t)}{\partial x^2}, \quad u(x,0) = \delta(x). \tag{25}$$

Here we show one case $p(\alpha)=2c\alpha$, then Eq. (25) is simplified as

$$\int_0^1 2ct^{1-\alpha} d\alpha \frac{\partial u(x,t)}{\partial t} = D \frac{\partial^2 u(x,t)}{\partial x^2}, \quad u(x,0) = \delta(x). \tag{26}$$

By using the identity,

$$\int_0^1 t^{1-\alpha} d\alpha = \frac{t-1}{\ln(t)}, \tag{27}$$

Eq. (26) is rewritten as

$$\frac{\partial u(x,t)}{\ln(t)/(2ct-2c)\partial t} = D \frac{\partial^2 u(x,t)}{\partial x^2}, \quad u(x,0) = \delta(x). \tag{28}$$

By using the identity,

$$\int \ln(t)/(t-1) dt = -\mathrm{dilog}(t), \tag{29}$$

and the structural derivative in Eq. (18) with the initial instance $t_0 = 0$, Eq. (28) is considered a structural diffusion equation .

$$\frac{\partial u(x,t)}{\partial (-\mathrm{dilog}(t)/2c)} = D \frac{\partial^2 u(x,t)}{\partial x^2}, \quad u(x,0) = \delta(x), \tag{30}$$

where dilog($t$) is the dilogarithm function [31],

$$\text{dilog}(t) = \int_1^x \frac{\ln(t)}{1-t} dt. \quad (31)$$

The corresponding mean squared displacement is

$$<x^2(t)> = -D\text{dilog}(t)/c. \quad (32)$$

In this case, the weight function determines the type of the underlying diffusion process.

## 3. Results

### 3.1 The evolution patterns in the distributed-order Hausdorff diffusion equation

To check the diffusion type of the distributed-order Hausdorff derivative model with varying values of $c$, Fig. 1 shows the mean squared displacements for the Fickian diffusion, the sub-diffusion with $\alpha = 0.8$, the logarithmic ultraslow diffusion with the structural function $k(t) = \ln(t)$ and the distributed-order Hausdorff derivative models with three different cases $c = 0.1$, 1.0 and 10. It can be observed from Fig. 1 that with the increasing time, the distributed-order Hausdorff derivative model can depict the super-diffusion when $c$ is 10, which diffuses a little faster than the Fickian diffusion. When $c$ is 1.0, the corresponding MSD grows faster than the logarithmic ultraslow diffusion, but slower than the classical sub-diffusion with power law $t^{0.8}$. In this case, the distributed-order Hausdorff derivative model characterizes the decelerating sub-diffusion process. Interestingly, it belongs to an ultraslow diffusion when $c$ is 0.1, where the MSD increases much slower than the logarithmic ultraslow diffusion. Thus, the distributed-order Hausdorff derivative model is flexible to describe the anomalous diffusion, i.e., both the super- and sub- diffusions, and the ultraslow diffusion.

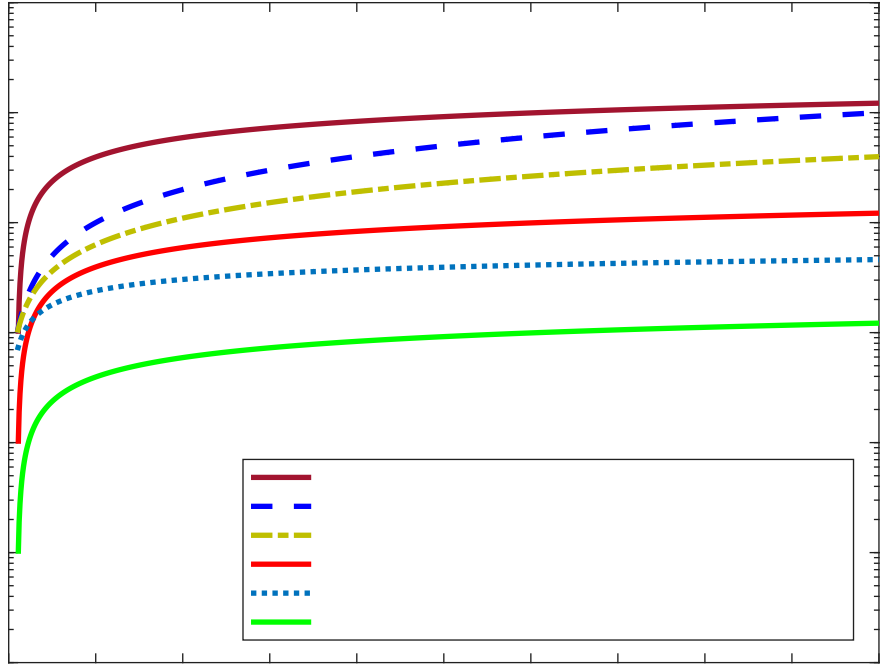

Fig. 1. Plots of the mean squared displacement for the Fickian diffusion, the sub-diffusion ($t^{0.8}$), the logarithmic ultraslow diffusion ($k(t) = \ln(t)$) and the distributed-order Hausdorff derivative models ($c = 0.1$, $1.0$ and $10$).

Fig. 2 illustrates the evolution patterns of the propagators for the traditional diffusion equation, Hausdorff derivative diffusion equation with $\alpha = 0.8$, the distributed-order Hausdorff derivative diffusion equation with $c = 1.0$ and the structural derivative diffusion equation with the kernel function $k(t) = \ln(t)$, where $D = 1$ and $x = 1$. As expected, in this case, the propagator of the decelerating sub-diffusion controlled by the distributed-order Hausdorff derivative diffusion equation decays more slowly than the sub-diffusion characterized by the time Hausdorff derivative diffusion equation, but decreases faster than the ultraslow diffusion controlled by the structural diffusion equation. The decay pattern in the decelerating sub-diffusion should be between the power law function in the sub-diffusion and the logarithmic function in the ultraslow diffusion for larger time. The results shown in Fig. 2 are consistent with those given in Fig. 1 in modeling the four types of diffusion processes.

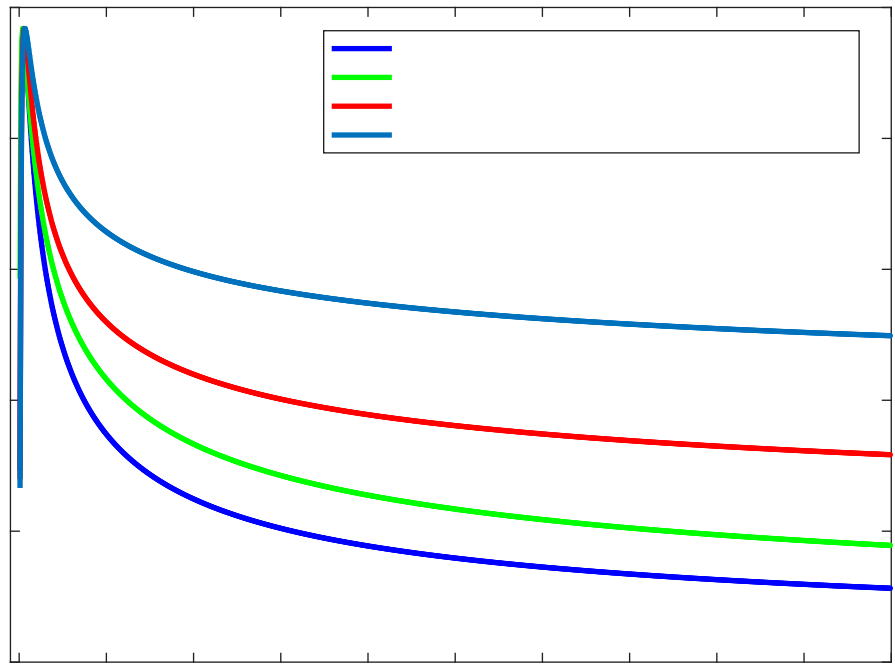

Fig. 2. The propagators of the traditional, Hausdorff ($\alpha = 0.8$), distributed-order Hausdorff ($c = 1.0$) and structural diffusion ($k(t) = \ln(t)$) equations from bottom to top versus $t$ for $D = 1$, $x = 1$.

## 3.2 Application cases to valid the distributed-order Hausdorff diffusion model

The solute diffusions in water-filled pore spaces of two-dimensional Euclidean and pore-solid prefractal porous media are investigated. The Euclidean and pore-solid prefractal porous media were respectively generated by the random allocation of pore cells with a solid initiator and the homogeneous algorithm with a pore-solid initiator [24]. The lattice sizes were set at 1000 by 1000 cells. Fig. 3 shows the relationship between the porosity $\phi$ and the lacunarity $L$, which is a quantitative measure of the spatial distribution of pores within the media. It satisfies a linear law in the Euclidean media $L \sim \phi$, but a power law function in the pore-solid prefractal media $L \sim \phi^{-0.89}$. The pore size distribution in the Euclidean porous media is uniform, and the pore-solid prefractal porous media contained both pore and particle cell size distributions. Solute diffusions in the above mentioned media was simulated with 100 particles using a stochastic cellular automaton with the myopic ant algorithm [24]. The diffusion time was set at 1000 incremental steps. More details on the simulated experiment can be found in [24].

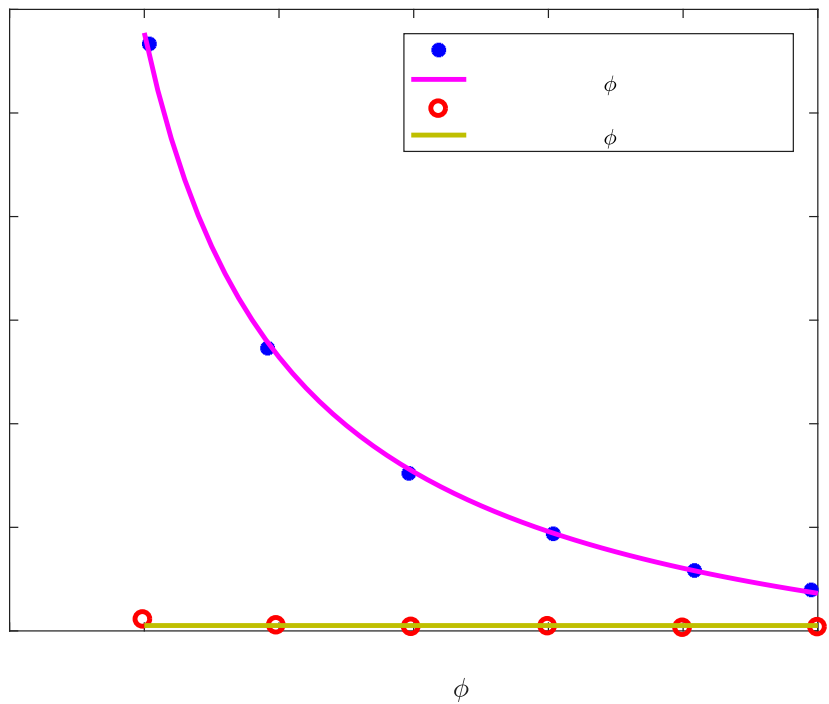

Fig. 3. The relationships between the lacunarity and the porosity for the Euclidean porous media ($L \sim \phi$) and the pore-solid prefractal porous media ($L \sim \phi^{-0.89}$) (based on data from [24]).

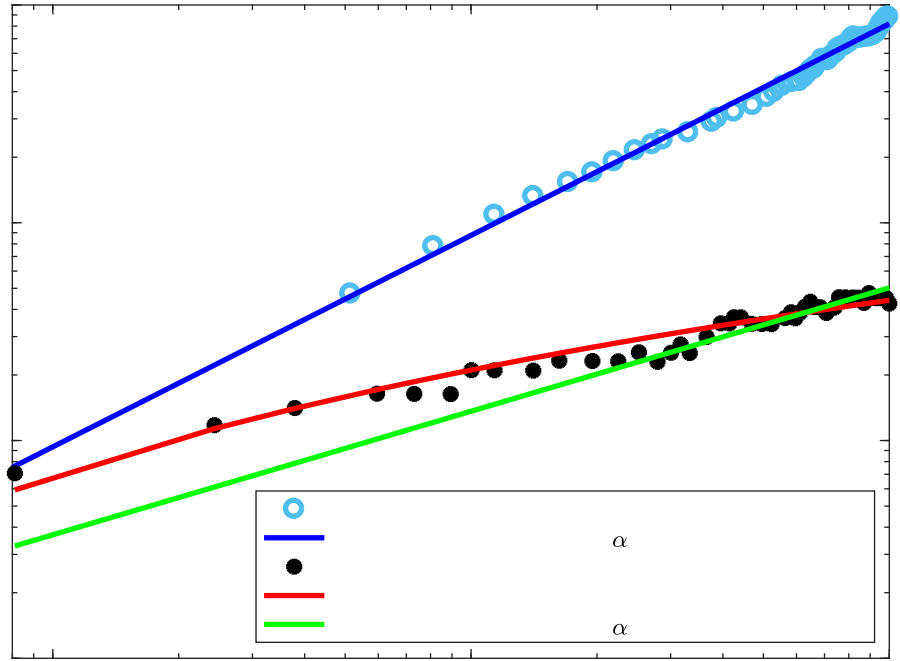

Fig. 4. The mean squared displacements of the tracer transport in the Euclidean and pore-solid prefractal porous media are respectively fitted by the distributed-order Hausdorff derivative model with $c = 1.73$ and the Hausdorff derivative model with $\alpha = 0.57$ (based on data from [24]).

Fig. 4 shows the mean squared displacement of the diffusive particles in both of the Euclidean and pore-solid prefractal porous media for a given porosity ($\phi = 0.5$). It is observed from Fig. 4 that the tracers in the pore-solid prefractal porous media diffuse faster than that in the Euclidean porous media, and both of the diffusion processes are not Fickian diffusion. From the viewpoint of geometrical feature, more large pores exist in the pore-solid prefractal porous media, and the Euclidean porous media has lager pore cluster size. However, the pores in the pore-solid prefractal porous media have more significant influence on the inside diffusion process compared with the pore cluster size. To differentiate the diffusion processes in both of the porous media and explain the corresponding physical mechanism, the distributed order Hausdorff derivative model and the Hausdorff derivative model with a constant order are used to describe the mean squared displacement, as illustrated in Fig. 4.

The MSD of the diffusion in the pore-solid prefractal porous media is almost a straight line on the double logarithmic axis, and can well be fitted by the Hausdorff derivative model with $\alpha = 0.97$. Thus, the tracer transport is a sub-diffusion in this pore-solid prefractal porous media. While it is a curve for the case of the Euclidean porous media, where the Hausdorff derivative model with $\alpha = 0.57$ only applies to the diffusion scenario at long time. It indicates that the MSD for the Euclidean porous media does not present mono-scaling. It is also noted that the tracers diffusive fast at short time, but become much slower at long time, which indicates that the motion is a decelerating sub-diffusion in the Euclidean porous media. We can observe from Fig. 4 that the distributed order Hausdorff derivative model with $c = 1.73$ is capable to capture all the features both at the short and long times. This distributed order Hausdorff derivative model generalizes the Hausdorff derive model and provides a more flexible tool to detect the tracer transport in porous media.

## 4. Discussion

This work provides a distributed order Hausdorff diffusion model to describe the tracer transport in porous media. The distribution of derivative order $p(\alpha)$ determines the type of the underlying diffusion process. In this study, the mean squared displacement of the tracer particles is derived in conjunction with the structural derivative when the order distribution $p(\alpha)=2c\alpha$, which can capture both accelerating and decelerating anomalous and ultraslow diffusions. The distributed order Hausdorff diffusion model is tested and effective to describe the tracer transport in water-filled pore spaces of two-dimensional Euclidean. It is demonstrated as a decelerating sub-diffusion, and can well be described by the Hausdorff diffusion model with a constant time derivative order. It should be pointed out that the tracer transport in the pore-solid prefractal porous media has only one scaling exponent, and is well fitted by the Hausdorff diffusion model with $\alpha = 0.97$. Compared with the time Hausdorff diffusion model, the distributed order Hausdorff diffusion model is more flexible to characterize the non-Fickian diffusion process in porous media.

It is noted that both of the distributed order time Hausdorff diffusion model and the time Hausdorff diffusion model have one variable parameter, which are respectively the parameter $c$ in the time order distribution and the time derivative order $\alpha$. The time derivative order $\alpha$ has clear physical meaning and is correlated with the diffusion coefficient of the porous media. The time Hausdorff diffusion model is equivalent with a diffusion equation with a nonlinear time dependent diffusion coefficient $D\alpha t^{\alpha-1}$. In the distributed order time Hausdorff diffusion model, the diffusion coefficient of the time derivative order can also be considered a function of the weight function $D / \int_0^1 p(\alpha)\alpha^{-1}t^{1-\alpha}d\alpha$. For the special case $p(\alpha)=2c\alpha$ used in this study, the diffusion coefficient is simplified as $D\ln(t)/(2ct-2c)$, and is independent with the derivative order. The heterogeneity in the porous media is respectively quantified and reflected by the derivative order α in the Hausdorff diffusion model and the parameter *c* in the distributed order Hausdorff diffusion model. The porous media we considered here are not real natural materials, thus some more field experiments of tracer transport in porous media are necessary to confirm the applicability of the model, including the concentration evolution in the porous media.

From the perspective of model selection, it is known that several models are successively provided to detect the diffusion processes in porous media with multi-scaling exponents, such as the distributed order fractional diffusion model, the variable orders Hausdorff and fractional diffusion models, and the tempered fractional diffusion model [32]. To select a reasonable model from the existing models for a specific scenario of tracer transport is an open issue. The physical mechanism and the application scope of each model are further needed to be summarized and explored. On the other hand, the parameters used to characterize the structure of the porous media, such as porosity, lacunarity and tortuosity, affect the solute transport process. Thus, the control equation of the tracer transport in the mentioned models should be improved by containing the mentioned parameters, and the quantitative relationships between the model parameters and the structural features of the porous media should be discussed in the further study.

## 5. Conclusions

This study introduces the distributed order Hausdorff derivative into the diffusion equation to detect the physical mechanism in multi-scaling non-Fickian diffusion, which is applied to investigate the tracer transport in heterogeneous porous media. Based on the foregoing results and discussion, the following conclusions are drawn:

1. In the distributed order time Hausdorff derivative diffusion model, when the weight function of distribution order is a linear function of the time derivative order $p(\alpha)=2c\alpha$, the MSD can be explicitly derived as a dilogarithm function in conjunction with the local structural derivative.

2. The distributed order time Hausdorff derivative diffusion model is equivalent with the diffusion equation with a nonlinear time dependent diffusion coefficient, which can capture both accelerating and decelerating anomalous and ultraslow diffusions by varying the values of parameter $c$ in the weight function.

3. The tracer transport in water-filled pore spaces of two-dimensional Euclidean porous media is demonstrated as a decelerating sub-diffusion, and can well be described by the distributed order Hausdorff diffusion model with $c = 1.73$, while the Hausdorff diffusion model with $\alpha = 0.97$ can fit the sub-diffusion of the tracer transport in the pore-solid prefractal porous media.

**Acknowledgments**

The work described in this paper was supported by the National Natural Science Foundation of China Nos. 11702085, 11572112, 41628202 and the Fundamental

**References**

[1] A. Zoia, C. Latrille, A. Cartalade, Nonlinear random-walk approach to concentration-dependent contaminant transport in porous media, Phys. Rev. E 79 (2009) 041125.

[2] Z. Dai, A. Wolfsberg, P. Reimus, et al, Identification of sorption processes and parameters for radionuclide transport in fractured rock, J. Hydrol. 414 (2012) 220-230.

[3] R. Chandrajith, S. Diyabalanage, K. Premathilake, et al, Controls of evaporative irrigation return flows in comparison to seawater intrusion in coastal karstic aquifers in northern Sri Lanka: evidence from solutes and stable isotopes, Sci. Total Environ. 548-549 (2016) 421-428.

[4] A. Werner, C. Simmons, Impact of sea-level rise on sea water intrusion in coastal aquifers, Groundwater, 47 (2010) 197-204.

[5] S. Baek, S. Kim, J. Kwon, et al, Ground penetrating radar for fracture mapping in underground hazardous waste disposal sites: A case study from an underground research tunnel, South Korea, J. Appl. Geophys. 141 (2017) 24-33.

[6] S. Lee, I. Yeo, K. Lee, et al, The role of eddies in solute transport and recovery in rock fractures: Implication for groundwater remediation, Hydrol. Process. 31 (20) 3580-3587 (2017).

[7] A. Einstein, On the movement of small particles suspended in stationary liquids required by molecular-kinetic theory of heat, Ann. Phys. 17 (1905) 549-560.

[8] R. Metzler, J. Jeon, A. Cherstvy, et al, Anomalous diffusion models and their properties: non-stationarity, nonergodicity, and ageing at the centenary of single particle tracking, Phys. Chem. Chem. Phys. 16 (2014) 24128-24164.

[9] R. Metzler, J. Klafter, The random walk's guide to anomalous diffusion: a fractional dynamics approach, Phys. Rep. 339 (2000) 1-77.

[10] I. Eliazar, J. Klafter, On the generation of anomalous and ultraslow diffusion, J. Physa-Math. Theor. 44 (2011) 2033-2039.

[11] W. Chen, Y. Liang, X. Hei, Structural derivative based on inverse Mittag-Leffler function for modeling ultraslow diffusion, Fract. Calc. Appl. Anal. 19 (2016) 1250-1261.

[12] S. Denisov, H. Kantz, Continuous-time random walk theory of superslow diffusion, Epl. 92 (2010) 2333-2358.

[13] A. Obembe, M. Hossain, S. Abu-Khamsin, Variable-order derivative time fractional diffusion model for heterogeneous porous media, J. Petrol. Sci Eng. 152 (2017) 391-405.

[14] X. Liu, H. Sun, M. Lazarevic, et al, A variable-order fractal derivative model for anomalous diffusion, Therm. Sci. 21 (2017) 51-59.

[15] Y. Zhang, C. Papelis, Particle-tracking simulation of fractional diffusion-reaction processes, Phys. Rev. E 84 (2011) 066704.

[16] W. Chen, Time-space fabric underlying anomalous diffusion, Chaos Soliton. Fract. 28 (2006) 923-929.

[17] H. Sun, Y. Zhang, W. Chen, et al, Use of a variable index fractional derivative model to capture transient dispersion in heterogeneous media, J. Contam. Hydrol. 157 (2014) 47-58.

[18] M. Meerschaert, E. Nane, P. Vellaisamy, Distributed-order fractional diffusions on bounded domains, J. Math. Anal. Appl. 379 (2011) 216-228.

[19] N. Su, P. Nelson, S. Connor, The distributed-order fractional diffusion-wave equation of groundwater flow: Theory and application to pumping and slug tests, J. Hydrol. 529 (2015) 1262-1273.

[20] Y. Liang, W. Chen, A non-local structural derivative model for characterization of ultraslow diffusion in dense colloids, Commun. Nonlinear Sci. 56 (2018) 131-137.

[21] H. Sun, M. Meerschaert, Y. Zhang, et al, A fractal Richards' equation to capture the non-Boltzmann scaling of water transport in unsaturated media, Adv. Water Resour. 52 (2013) 292-295.

[22] W. Chen, F. Wang, B. Zheng, et al, Non-Euclidean distance fundamental solution of Hausdorff derivative partial differential equations, Eng. Anal. Bound. Elem. 84 (2017) 213-219.

[23] Y. Liang, A. Ye, W. Chen, et al, A fractal derivative model for the characterization of anomalous diffusion in magnetic resonance imaging, Commun. Nonlinear Sci. 39 (2016 ) 529-537.

[24] J. Kim, E. Perfect, H. Choi, Anomalous diffusion in two-dimensional Euclidean and prefractal geometrical models of heterogeneous porous media, Water Resour. Res. 43 (2017) W01405.

[25] M. Hamani, Nonlinear boundary value problems for differential inclusions with Caputo fractional derivative, J. Juliusz Schauder Cent. 32 (2008) 115-130.

[26] F. Mainardi, A. Mura, G. Pagnini, et al, Time-fractional diffusion of distributed order, J. Vib. Control 14 (2008) 1267-1290.

[27] A. Chechkin, R. Gorenflo, I. Sokolov, Retarding subdiffusion and accelerating superdiffusion governed by distributed-order fractional diffusion equations, Phys. Rev. E 66 (2002) 046129.

[28] W. Chen, Non-power-function metric: a generalized fractal, Math. Phys. viXra: (2017) 1612.0409.

[29] Y. Sinai, The limiting behavior of a one-dimensional random walk in a random medium, Theor. Probab. Appl. 27 (1983) 256-268.

[30] W. Cai, W. Chen, W. Xu, Characterizing the creep of viscoelastic materials by fractal derivative models, Int. J. Non-Linear Mech. 87 (2016) 58-63.

[31] D. Zagier, The Dilogarithm Function, Publ. Math. 89 (2007) 321-330.

[32] Y. Zhang, Moments for Tempered Fractional Advection-Diffusion Equations, J. Stat. Phys. 139 (2010) 915-939.